\documentclass[twocolumn,aps,prl,nopacs,floatfix,longbibliography]{revtex4-2}
\usepackage{graphicx}
\usepackage{epsfig}
\usepackage{rotating}
\usepackage{amsmath}
\usepackage{amsfonts}
\usepackage{amssymb}
\usepackage{enumerate}
\usepackage{longtable}
\usepackage{dcolumn}
\newcolumntype{.}{D{.}{.}{-1}}
\usepackage{tabularx}
\usepackage{dcolumn}% Align table columns on decimal point
\usepackage{bm}

\usepackage[]{hyperref}
\hypersetup{colorlinks=true,linkcolor=blue,citecolor=blue,urlcolor=blue,pdfpagemode=UseNone}
\DeclareMathOperator{\Tr}{Tr}

\begin{document}

\title{Quantum Criticality in the infinite-range Transverse Field Ising Model}

\author{Nicholas Curro, Kaeshav Danesh and Rajiv R. P. Singh}
\affiliation{University of California Davis, CA 95616, USA}

\date{\rm\today}

\begin{abstract}
We study quantum criticality in the infinite range Transverse-Field Ising Model.
We find subtle differences with respect to the well-known single-site mean-field theory, especially in terms of
gap, entanglement and quantum criticality. The calculations are based on numerical diagonalization of Hamiltonians
with up to a few thousand spins. This is made possible by the enhanced symmetries of the model, which divide the
Hamiltonian into many
%-many 
block-diagonal sectors. The finite temperature
phase diagram and the characteristic jump in heat capacity closely resemble the behavior in mean-field theory. 
However, unlike mean-field theory where excitations are always gapped, the excitation gap in the infinite range
model goes to zero from both the paramagnetic side
and from the ferromagnetic side on approach to the quantum critical point. Also, contrary to mean-field theory, 
at the quantum critical point the Quantum Fisher
Information becomes large, implying long-range multi-partite entanglement. We find that the main role of temperature is to shift statistical weights from one conserved sector to another. However, low energy excitations in each sector arise only near the quantum critical point implying that low energy quantum fluctuations can  arise only in the vicinity of the quantum critical field where they can persist up to temperatures of order the exchange constant.
\end{abstract}

%\pacs{74.70.-b,75.10.Jm,75.40.Gb,75.30.Ds}

\maketitle

\section{Introduction}
\label{sec:intro}
Recent studies of quadrupolar transverse-field Ising behavior in thulium vanadate materials  have raised important questions about quantum criticality in such systems \cite{Massat2021,Vinograd2022,Wang2021}. The finite temperature thermodynamic behavior shows a jump in the heat-capacity characteristic of mean-field behavior that is expected for phonon-mediated systems with long-range interactions \cite{Massat2021, Gehring1975}. Yet, surprisingly, NMR spin-echo experiments show  a dramatic wipe-out phenomena extending over a fan-like region in the temperature-transverse field plane near the quantum critical field, signifying persistent quantum critical fluctuations \cite{Nian2024}. The well known single-site mean-field theory captures the phase diagram and the jump in the heat capacity quite well. However, the mean-field theory always has a large energy gap and hence no low energy fluctuations that can cause rapid decoherence of NMR signals \cite{Chen2013}. The purpose of this work is to study a concrete model that has mean-field thermodynamic behavior and yet may support persistent quantum critical fluctuations.

We consider a system of $N$ half integer spins.
The infinite range Transverse-Field Ising Model (TFIM) Hamiltonian can be expressed in terms of Pauli spin matrices 
for the $i$th spin $\sigma_i^\alpha$ as
\begin{equation}
    {\cal H}= - \frac{J}{2N} (\sum_i \sigma_i^z)^2 - h_x \sum_i \sigma_i^x.
\end{equation}
We will set $J=1$, which sets all energy scales.

\section{Mean-field theory}
\label{sec:mft}
 The standard mean-field theory replaces the Hamiltonian by a single site effective Hamiltonian \cite{deGennes,Stinchcombe}
 \begin{equation}
    {\cal H}_{eff}= - m_z \sigma^z - h_x \sigma^x,
\end{equation} 
where the magnetization $m_z$ needs to be determined self-consistently
\begin{equation}
    m_z = \langle \sigma^z \rangle.
\end{equation}
Various physical properties are easy to calculate in this approximation. It leads to a transition temperature as a function of transverse-field given by
\begin{equation}
    T_c(h_x) = h_x/\tanh^{-1}{(h_x)}.
\end{equation}
In our units the zero field transition temperature $T_c(0)$ and the critical field at zero temperature $h_c$ are both unity.

The mean-field approximation is known to be exact in the thermodynamic limit for the classical model ($h_x=0$). In this study we explore deviations from the mean-field behavior for the infinite-range model for non-zero $h_x$ and the implications for quantum critical phenomena. We note that the infinite range model has been derived as a zeroth order model for the thulium vanadate and other quadrupolar transverse-field Ising materials where gapless phonons mediate long-range interactions between the Ising degrees of freedom \cite{Gehring1975,Schmalian,Paul,Massat2021}.

\section{Infinite-range model}
\label{sec:symmetries}
The infinite range model has a large number of symmetries. Because the Hamiltonian depends only on two operators
\begin{equation}
    S_z= \sum_i S_i^z,
\end{equation}
and
\begin{equation}
    S_x= \sum_i S_i^x,
\end{equation}
both of which commute with the total spin operator
\begin{equation}
    S_t^2= \sum_i S_i^2,
\end{equation}
the Hamiltonian becomes block diagonal into many relatively small ($O(N)$) Hilbert-space sectors \cite{Zoller,Wilms}.
The energy levels and properties in each sector depend on the total spin $s_t$, which can take values from $0$ to $N/2$ (we will assume $N$ is even). The spin sector has Hilbert space dimension $2s_t+1$. The number of copies of the spin $s_t$ sector is the number of ways $N$ spin-half objects can be combined into total spin $s_t$. This number $d(s_t)$ is one for $s_t=N/2$ and for $s_t<N/2$ can be expressed in terms of the combinatoric factors $C^m_n$ as
\begin{equation}
    d(s_t)= C^{N}_{N/2+s_t} - C^{N}_{N/2+s_t+1}.
\end{equation}
In each $s_t$ sector the Hamiltonian matrix can be written out in terms of the matrix elements of spin operators and diagonalized numerically. Thus, given the symmetries it is possible to numerically calculate energy eigenlevels of the system with up to a few thousand spins rather than few tens of spins for most lattice models. 

\subsection{Energy gap}
\label{sec:gap}. 

\begin{figure}
\begin{center}
\includegraphics[width=0.8\columnwidth]{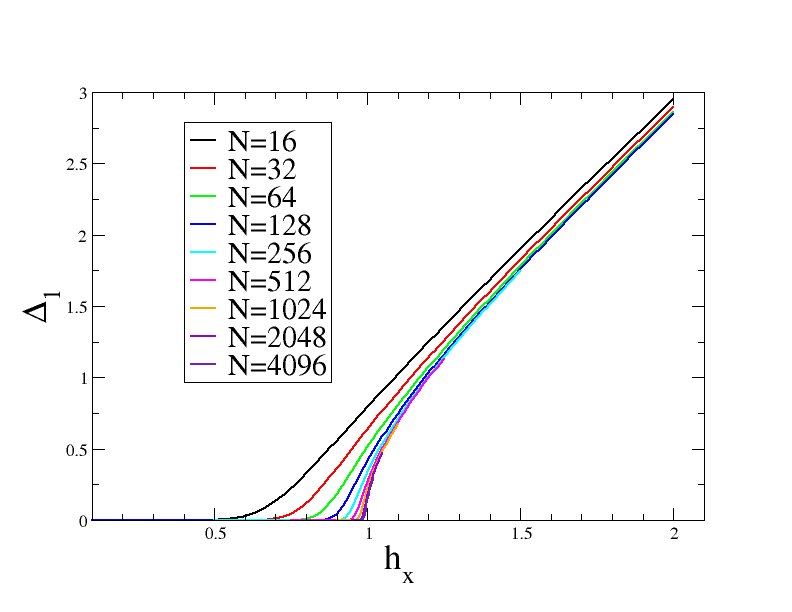}
\caption{
Energy difference between the lowest two states. Note that for $h<h_c=1$, there are two degenerate states in the thermodynamic limit and hence this energy difference goes to zero. Mean-field theory results are shown by dashed lines.
\label{Delta1}}
\end{center}
\end{figure}

\begin{figure}
\begin{center}
\includegraphics[width=0.8\columnwidth]{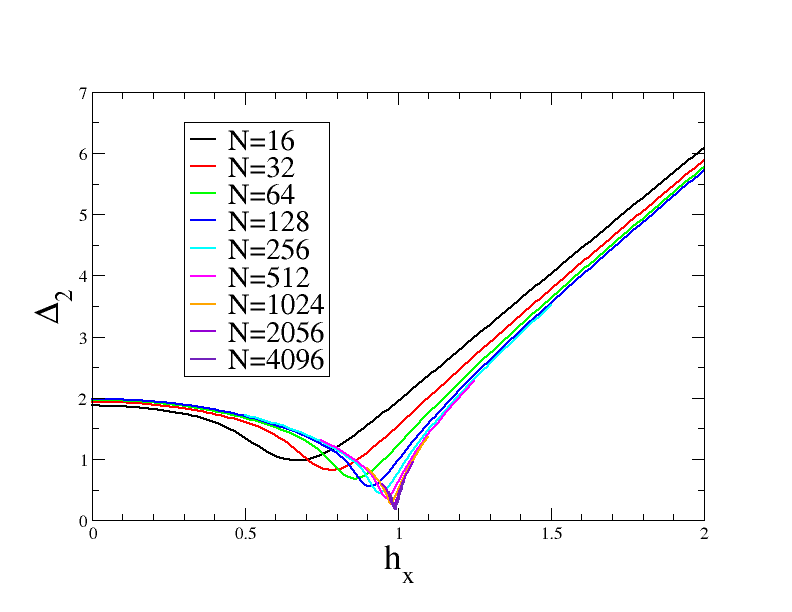}
\caption{
Energy difference between the third lowest energy state and the ground state showing that excitation energy goes to zero from
both sides as one approaches the quantum critical point. Mean-field theory results are shown by dashed lines.
\label{Delta2}}
\end{center}
\end{figure}

The ground state always lies in the largest $s_t$ sector. Energy gap between the lowest excited states and the ground state are shown in Fig.~1 and Fig.~2 as a function of the transverse-field $h_x$. In the $N\to \infty$ limit, there are two degenerate ground states in the ordered phase at $h_x<1$. One can see the difference between the energies of the first two states in Fig.~1. With increasing system size this is rapidly going to zero for $h<h_c$. To measure the excitation energy below $h_c$ one needs to look at the next excited state in the ordered phase. This excitation energy is shown in Fig.~2.

From the two figures, a key difference from the single site mean-field theory becomes clear. In mean-field theory
the gap does not go below $2J$. 
%\textcolor{red}{(maybe we could include a dotted line showing the mean field result?) }
It goes to zero at $h_c$ in the infinite-range model as expected at the quantum critical point. The gap vanishes as one approaches the quantum critical point $h_c$ from either side. Further more, at the quantum critical point, one expects a whole cascade of states to come down to zero energy in the thermodynamic limit. 
%We can see two of them doing so from the high-field side in the figures.

To see how the gap vanishes with system size $N$ at the quantum critical point, we show a log-log plot of several energy gaps as a function of system size in Fig.~3. All the states shown exhibit gaps vanishing as $N^{-1/3}$ as $N\to\infty$ (See also Ref.~\cite{Dusuel}).

\begin{figure}
\begin{center}
\includegraphics[width=0.98\columnwidth]{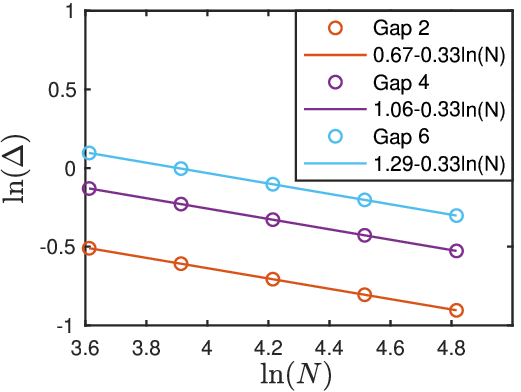}
\caption{Log-log plot of energy gaps vs $N$ at the quantum critical point for several low lying states showing the gap scaling as $N^{-1/3}$.
 \label{gap-vs-N}}
\end{center}
\end{figure}

%The comparison of the lowest 2 energy gaps in a given $s_t$ sector with mean-field theory is shown in Fig.~3 up to high values of the field. 
The two lowest excitations in the paramagnetic phase, within a sector, can be identified with one and two spin-flip excitations. Even for large $h_x$ the gap stays below the mean-field value by an amount $J=1$ and $2J=2$ within a given $s_t$ sector. As shown in the Appendix, this result can be obtained from perturbation theory around the large field limit and shows that deviations from mean-field theory are present for all fields and not just near the critical field.

% In Fig.~3, we show the spectral function associated with the total transverse spin operator. In these finite systems, the spectral weights consist of a series of delta-function peaks that have been broadened
% by a Lorentzian. One can see that away from the quantum critical point, there is a large gap in the spectra which is almost size
% independent. In contrast, at the critical point a whole series of states are converging towards zero energy. It is this continuum at
% low energies, characteristic of a quantum critical point, that can cause a rapid decoherence of the nuclear spins.

\subsection{Quantum Fisher Information}
\label{sec:qfi}

\begin{figure}
\begin{center}
\includegraphics[angle=270,width=0.8\columnwidth]{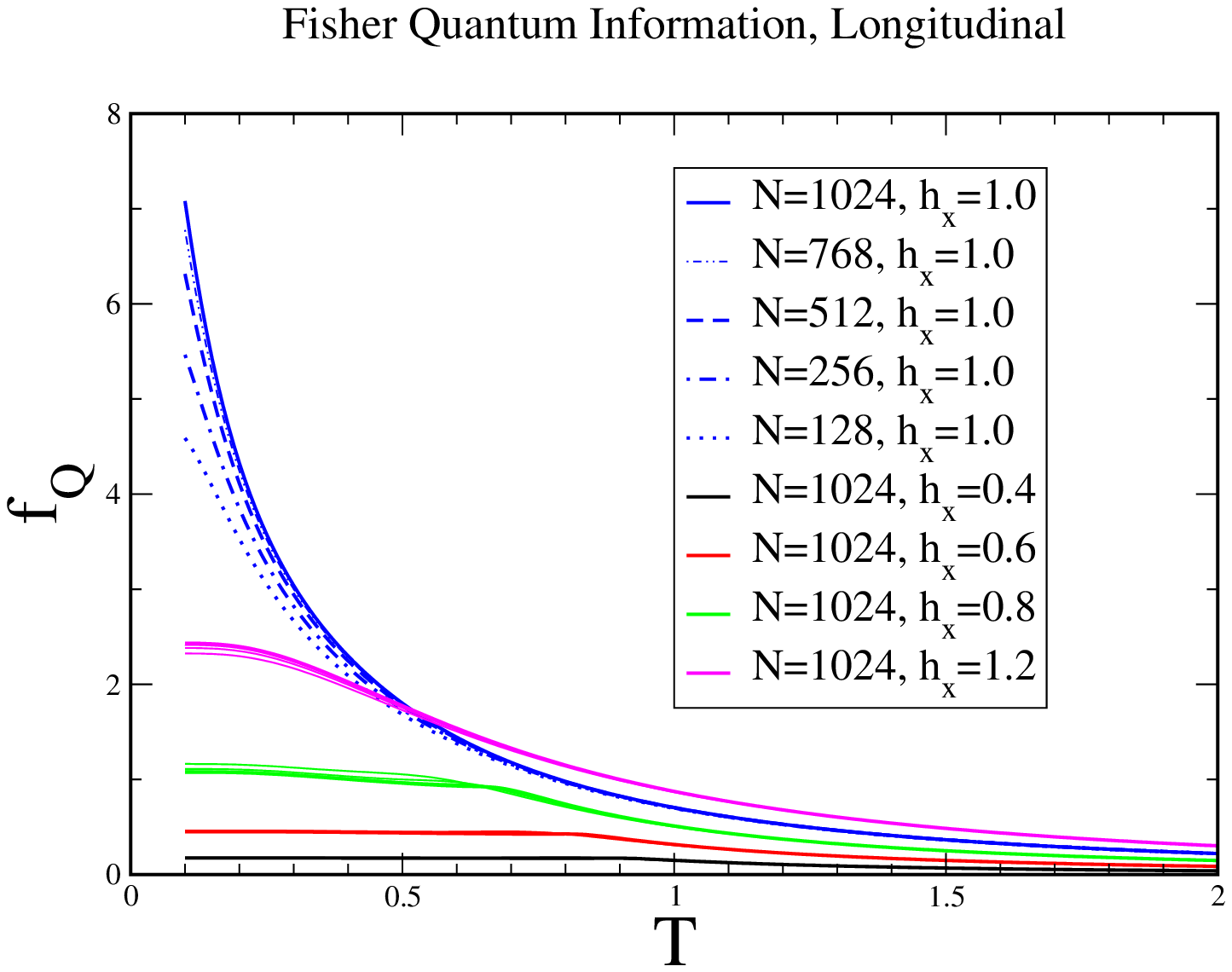}
\caption{
A plot of the Quantum Fisher Information further confirms the quantum critical nature by showing the build up of quantum entanglement in the thermodynamic state at low temperatures near the quantum critical point.
\label{QF}}
\end{center}
\end{figure}

\begin{figure}
\begin{center}
\includegraphics[width=0.8\columnwidth]{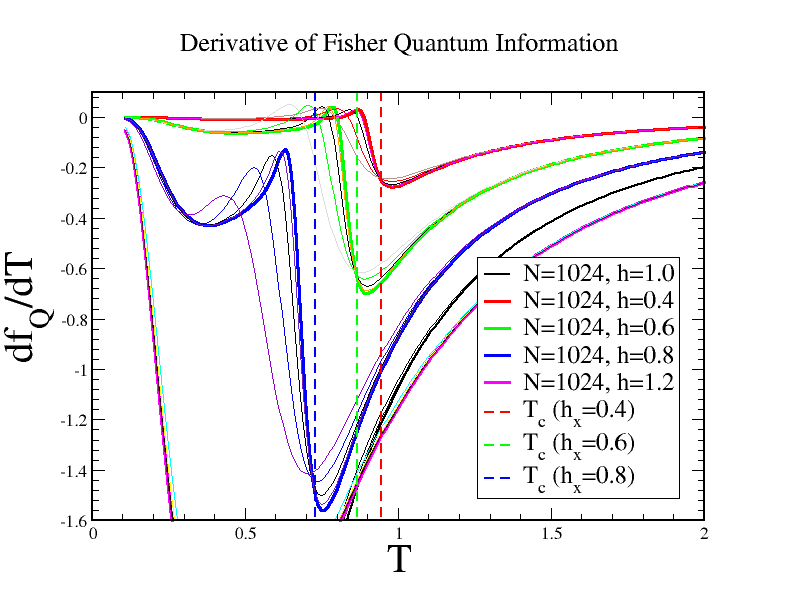}
\caption{
Derivative of the Quantum Fisher Information with respect to temperature shows that $f_Q$ inherits the energy singularity and hence its derivative has a discontinuity like the heat capacity at the finite temperature transition. Note that the thinner lines correspond to the same parameter as the solid line but for smaller system sizes ($N=768$, $512$ and $256$)to indicate how the curves move with size of the system.
\label{QF-deriv}}
\end{center}
\end{figure}

In mean field theory, the quantum state factorizes and does not have any quantum entanglement between different sites,
even at the quantum critical point. One can consider multi-particle entanglement as one of the defining properties of quantum criticality. To look for multi-partite entanglement in the infinite-range model, 
we turn to Quantum Fisher Information QFI \cite{Zoller}. QFI is the generalization
of Fisher Information from classical statistics to a quantum mechanical system. 
%It asks the question: 
%How much does an observable
%discriminate between different parameters that define the wave-function or the density matrix of the system. In other words,
%how well does measuring one observable constrain the parameter and the phase of the system. 
For a pure state, the QFI associated with a variable $\hat O$ is simply proportional to its variance. 
%Thus for an extensive variable like magnetization, QFI is its structure factor, a well studied property in Condensed Matter Physics to inform on
%the state of the system. 
In a mixed state, such as at finite temperatures, let the $i$th eigenstate of the system have probability $p_i$. Then the QFI
%Quantum Fisher Information 
$F_Q$ for the observable $\hat O$ is given by the expression \cite{Zoller}:
\begin{equation}
    F_Q= 2\sum_{i,j} \frac{(p_i-p_j)^2}{p_i+p_j} |<i|\hat O|j>|^2.
\end{equation}
%This quantity can be expressed as an integral of the dynamical susceptibility with an appropriate integrating factor.
%One can then extremize over all possible extensive observables to call it the Quantum Fisher Information 
%associated with the thermodynamic state.

The QFI is a witness of multi-particle entanglement. For a system of $N$
spin-half objects, if the QFI per site for any extensive variable $\hat O$ defined as
\begin{equation}
    f_Q= F_Q/N,
\end{equation}
exceeds some integer $m$ (for sufficiently large $N$), then there is at least $m+1$-particle entanglement in the
system. 
%No state with at most $m$ particle entanglement can have such a QFI. 
%Note, however, that a large Quantum Fisher Information is a sufficient condition for multi-partite entanglement. It is not a necessary condition.

The operator with the largest Quantum Fisher Information for our model is:
\begin{equation}
    \hat{O}=S_z,
\end{equation}
where $S_z$ is the total z-component of the spin-operator defined in Eq.~5. All QFI calculations will be shown for this operator.
Figure 4 shows $f_Q$ for several system sizes. Note that for
a proper calculation at finite temperatures one must sum over all states of the system and not just consider one spin
sector of the Hilbert space \cite{Zoller}. It is clear
that near the quantum critical point at low temperatures $f_Q$ becomes large and this means that the system becomes highly
entangled.
Note also that the Quantum Fisher Information decreases more rapidly at $T=0$ in the ordered phase than in the disordered phase. This is consistent with the behavior of other measures of quantum entanglement such as Renyi entanglement entropy (See e.g. Ref.~\cite{Melko}).
%and thus the infinite-range model genuinely supports a quantum critical point in its phase diagram contrary to mean-field theory.

Figure 5 shows the derivative of $f_Q$ with respect to temperature for different transverse field values. One finds a characteristic behavior of a sharp maxima followed by a minima, which presumably becomes a discontinuity in the thermodynamic limit. This feature tracks the critical temperatures at different fields which are shown by dashed lines. This is not surprising as measures of quantum entanglement can inherit an energy-like singularity even at a finite temperature phase transition, just from the singular changes in the density of states at the finite temperature critical point \cite{Grover1,Grover2,Sherman}. Note however that this finite temperature singularity does not mean there is any enhanced multi-particle quantum entanglement at the classical phase transition, as the magnitude of $f_Q$ remains small away from the quantum critical region.

\begin{figure}
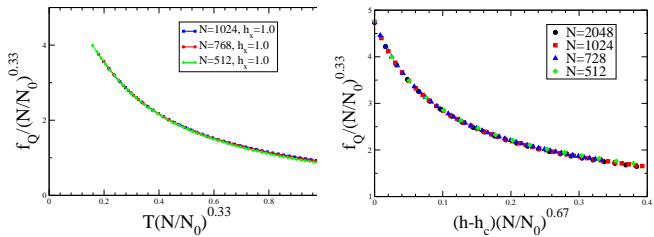

\begin{center}
\includegraphics[width=0.49\columnwidth]{fqi-l-hc.eps}
\includegraphics[width=0.49\columnwidth]{qfi-T0.eps}
\caption{
Scaling plots of Quantum Fisher Information $f_Q$ at critical field $h_c$ versus $T$ (left) and at $T=0$ versus transverse-field (right) for different system sizes $N$. We have scaled $N$ by a fixed reference size $N_0=128$.
\label{QFI-crit}}
\end{center}
\end{figure}

\begin{figure}
\begin{center}
\includegraphics[width=0.49\columnwidth]{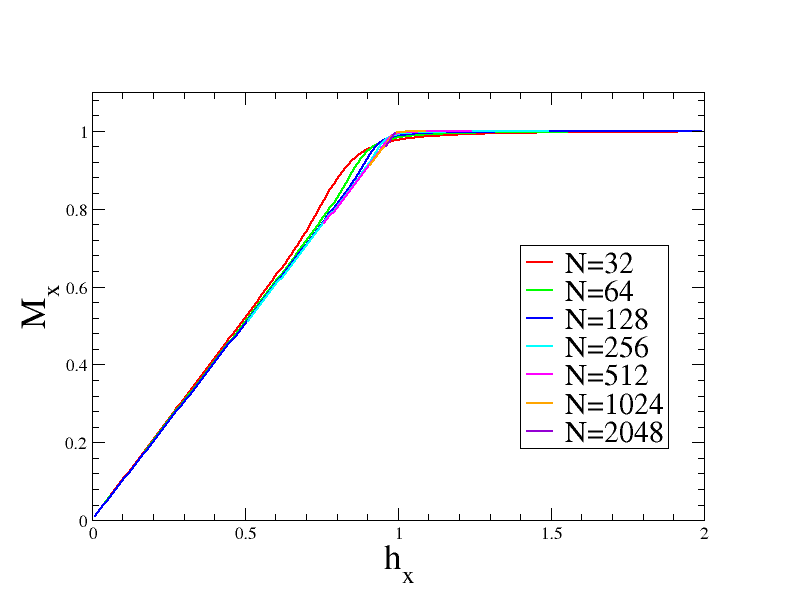}
\includegraphics[width=0.49\columnwidth]{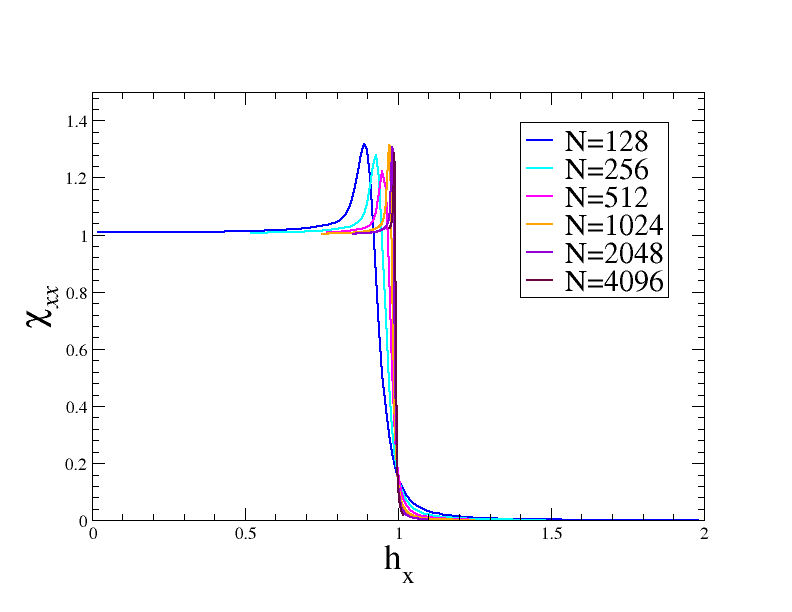}
\caption{
Plot of the transverse magnetization $M_x$ (left) and transverse susceptibility $\chi_{xx}$ (right) as a function of the transverse field in the ground state of the system. Mean-field results are shown by dashed lines.
\label{Mx}}
\end{center}
\end{figure}

\begin{figure}
\begin{center}
\includegraphics[width=0.49\columnwidth]{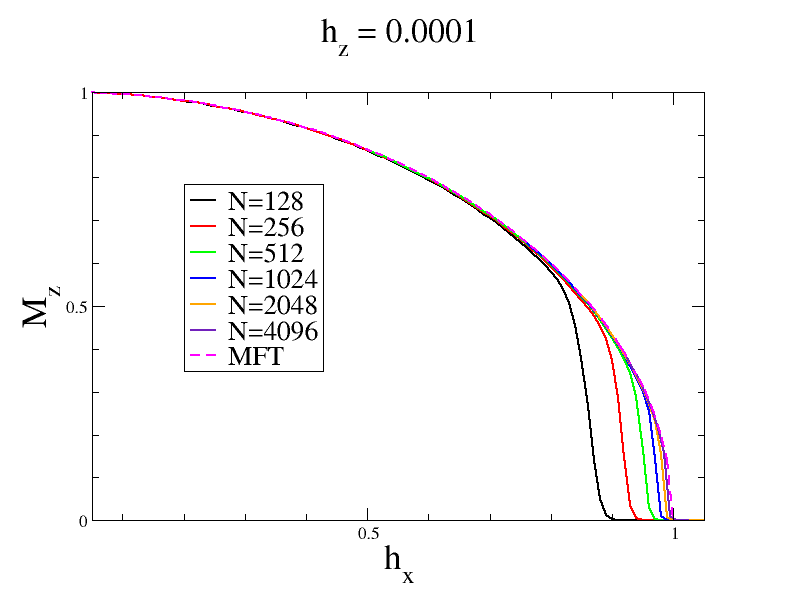}
\includegraphics[width=0.49\columnwidth]{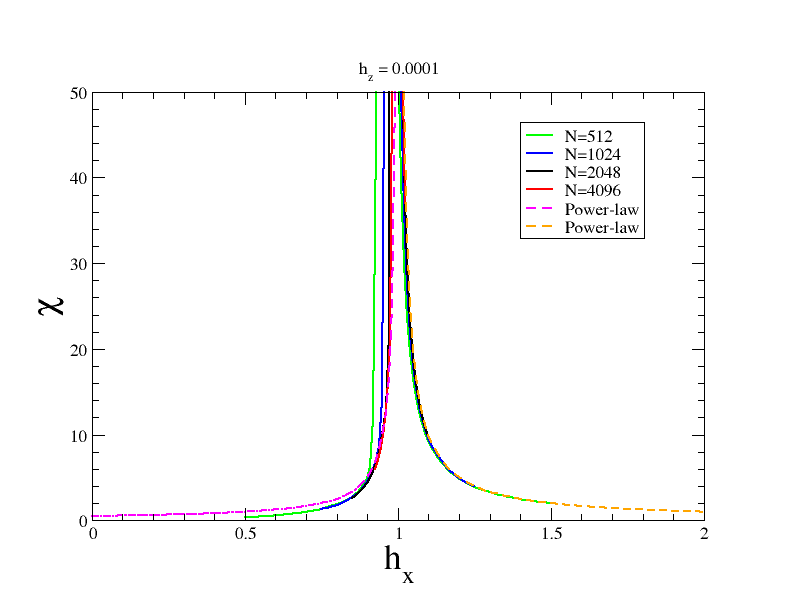}
\caption{
A plot of the longitudinal magnetization (left) and longitudinal susceptibility (right) as a function of the transverse-field in the ground state of the system obtained with a small symmetry-breaking field
$h_z=0.0001$.
\label{Mz}}
\end{center}
\end{figure}

% \begin{figure}
% \begin{center}
% \includegraphics[width=0.8\columnwidth]{chixx.eps}
% \caption{
% A plot of the transverse susceptibility in the ground state of the system as a function of the transverse field.
% \label{Chixx}}
% \end{center}
% \end{figure}

% \begin{figure}
% \begin{center}
% \includegraphics[width=0.8\columnwidth]{chiz.eps}
% \caption{
% A plot of the longitudinal susceptibility as a function of the transverse field in the ground state of the system, which diverges strongly at the quantum critical
% point.
% \label{Chiz}}
% \end{center}
% \end{figure}

\subsection{Quantum Critical Properties of the Quantum Fisher Information}

At $T=0$, QFI reduces to the variance of the operator $S_z$ in the ground state of the system, up to an overall factor. Near the quantum critical point $N\to\infty$,
$T\to 0$, $h\to h_c$, homogeneity \cite{Widom} implies, we can write the singular piece of QFI as
\begin{equation}
    f_Q(N,T,h)= N^a X(T\ N^b, (h-h_c)\ N^c),
\end{equation}
where $X(x,y)$ is a scaling function of two variables.
Fig.~6 shows the scaling plots for (a) $h=h_c$ and (b) $T=0$ with exponents $a=1/3$, $b=1/3$, $c=2/3$. Note that the exponent $b$ is consistent with the expected $T/\Delta$ scaling with the gap $\Delta$ vanishing with power $N^{-1/3}$.

Our scaling relations imply that in the thermodynamic limit, at $h=h_c$, $f_Q$ diverges as $1/T$ as $T\to 0$ and furthermore, at $T=0$, $f_Q$ diverges as $(h-h_c)^{-1/2}$. The latter exponent is consistent with the expected divergence in the mean-field universality class with a critical exponent $\gamma-z\nu=1/2$, with $\gamma=1$, $z=1$ and $\nu=1/2$.

\subsection{Ground-state magnetization and susceptibilities}
\label{sec:mx}
The transverse magnetization $M_x$ and susceptibility $\chi_{xx}$ in the ground state can be obtained by taking derivatives of the ground state energy with $h_x$. To calculate the longitudinal magnetization or order parameter $M_z$ and the order parameter susceptibility $\chi_{zz}$ one needs to add a small
symmetry-breaking field. These quantities are shown in Figs.~7 and 8. %\textcolor{red}{It might be worthwhile to stack figs 5 and 6 together with the same x-axis, and the same for Figs. 7-8.  Also, maybe we could include a dotted line showing MF behavior.} 
Deviations from mean-field theory is most pronounced in the transverse susceptibility near the quantum critical point $h_x=h_c$. In the mean-field theory the transverse susceptibility is a constant below $h_c$ and the magnetization
$M_x$ increases linearly with the field. In the infinite range model there is very little deviation from mean-field results in the thermodynamic limit ($N\to\infty$) away from the critical point. However, an enhanced susceptibility very near the critical
point with a sharp peak at the transition seems to persist with increasing $N$ as seen in Fig.~7. Note that the jump in the transverse susceptibility reflects the mean-field critical behavior and is not special to the infinite-range model.
%showing further evidence for quantum critical fluctuations. 
Singularities in longitudinal magnetization and susceptibility are consistent with mean-field critical exponents as shown by fits to mean-field behavior in Fig 8. That is,
$M_z\simeq |h_x-h_c|^\beta$ and $\chi_{zz}\simeq |h_x-h_c|^{-\gamma}$ with
$\beta=1/2$ and $\gamma=1$.
%It is quite plausible that in long-range models, where the range is long but not infinite, and the system shows mean-field behavior at the transition, similar enhancements near the quantum critical point arise.
%It should be possible to observe these peaks in experiments in systems that are well modelled with long-range interactions.

\section{Finite Temperature Thermodynamic Properties and Thulium Vanadates}
\label{sec:thermo} 

\begin{figure}
\begin{center}
\includegraphics[width=0.8\columnwidth]{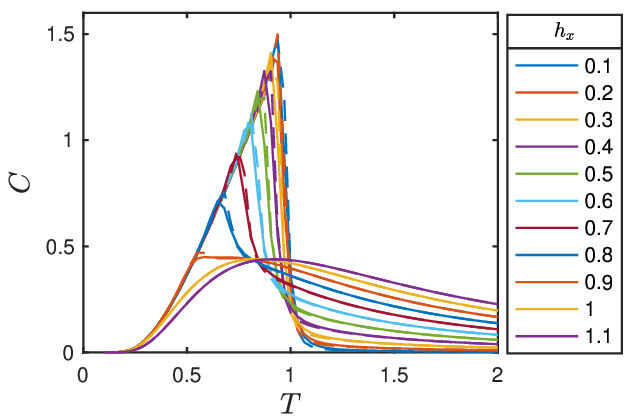}
\caption{
Heat capacity as a function of temperature for various values of the transverse field for $N=1024$ spins. The solid lines are for $N=1024$ and the dashed lines for $N=2048$. 
\label{Cv}}
\end{center}
\end{figure}

%\begin{figure}
%\begin{center}
%\includegraphics[width=0.8\columnwidth]{N512-chi-T.png}
%\caption{
%A plot of the transverse susceptibility as a function of temperature for fields ranging from $h_x=0.1$ to $h_x=2.1$
%\label{fig}}
%\end{center}
%\end{figure}

\begin{figure}
\begin{center}
\includegraphics[width=0.8\columnwidth]{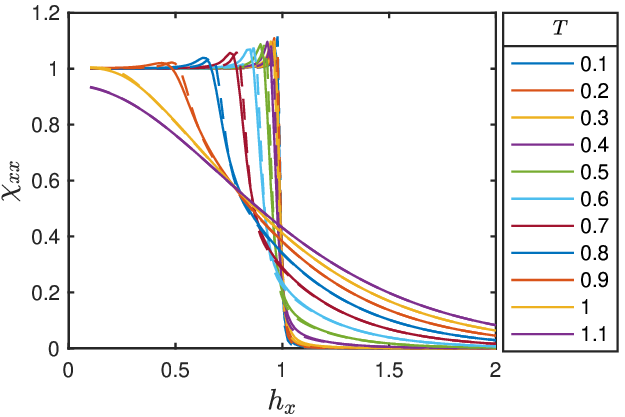}
\caption{
Transverse susceptibility as a function of transverse-field for fields for various temperatures for the infinite-range model.
Solid lines are for $N=1024$ and dashed lines for $N=2048$.
\label{Chixvshx}}
\end{center}
\end{figure}

\begin{figure}
\begin{center}
\includegraphics[width=0.8\columnwidth]{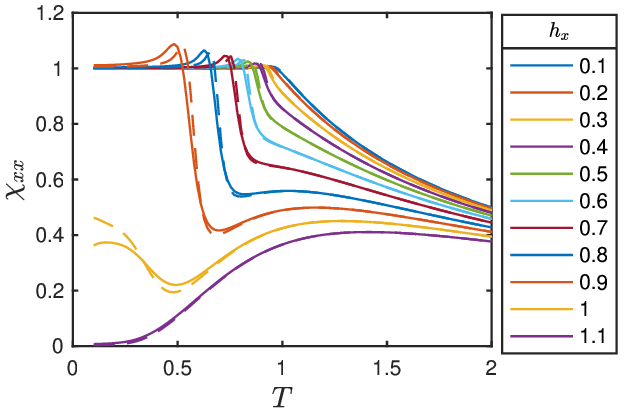}
\caption{
Transverse susceptibility as a function of temperature for various transverse fields for the infinite-range model.
Solid lines are for $N=1024$ and dashed lines for $N=2048$.
\label{ChixvsT}}
\end{center}
\end{figure}

%\section{Heat capacity/Entropy at low T}
%\label{sec:Cv}. 
%Heat capacity at low temperatures

Thermodynamic properties such as entropy, heat capacity and transverse susceptibilities at finite temperatures 
can be obtained from the
partition function of the system. The partition function is given by: 
\begin{eqnarray}
    Z(T,h_x)& =& \Tr \exp(-\beta\mathcal{H}) \nonumber \\
         &=& \sum_{s_t} d(s_t) \sum_{i\in s_t} e^{-\beta E_i},
\end{eqnarray}
where $d(s_t)$ is the number of sectors with spin $s_t$ in the system.

A plot of the calculated heat capacity as a function of temperature is shown in Fig.~9 for a range of 
transverse-field values. The system studied is large enough to exhibit the characteristic jump like feature of
mean-field theory as observed in the TmVO$_4$ materials \cite{Massat2021}.

%\section{Transverse-susceptibility}
%\label{sec:chi}. 

%The transverse-susceptibility as a function of temperature for various field values is shown in Figure 10.
Transverse-susceptibility as a function of field for various temperatures is shown in Figure 10, and 
%transverse-susceptibility
as a function of temperature for various fields  in Figure 11.
%The corresponding plots for the mean-field theory are also shown for comparison. 
In mean-field theory the susceptibility is constant in the ordered phase and jumps at the transition as a function of field and then rapidly
decreases in the paramagnetic phase.
The most visible
difference from mean-field theory is the rise of the transverse susceptibility in the ordered phase above the mean-field value and a small
but distinct peak near the quantum critical point. At low temperatures, this peak increases with increase in temperature. Such a
peak may be observable in the materials.

\subsection{The TmVO$_4$ materials}

\begin{figure}
\includegraphics[width=0.9\columnwidth]{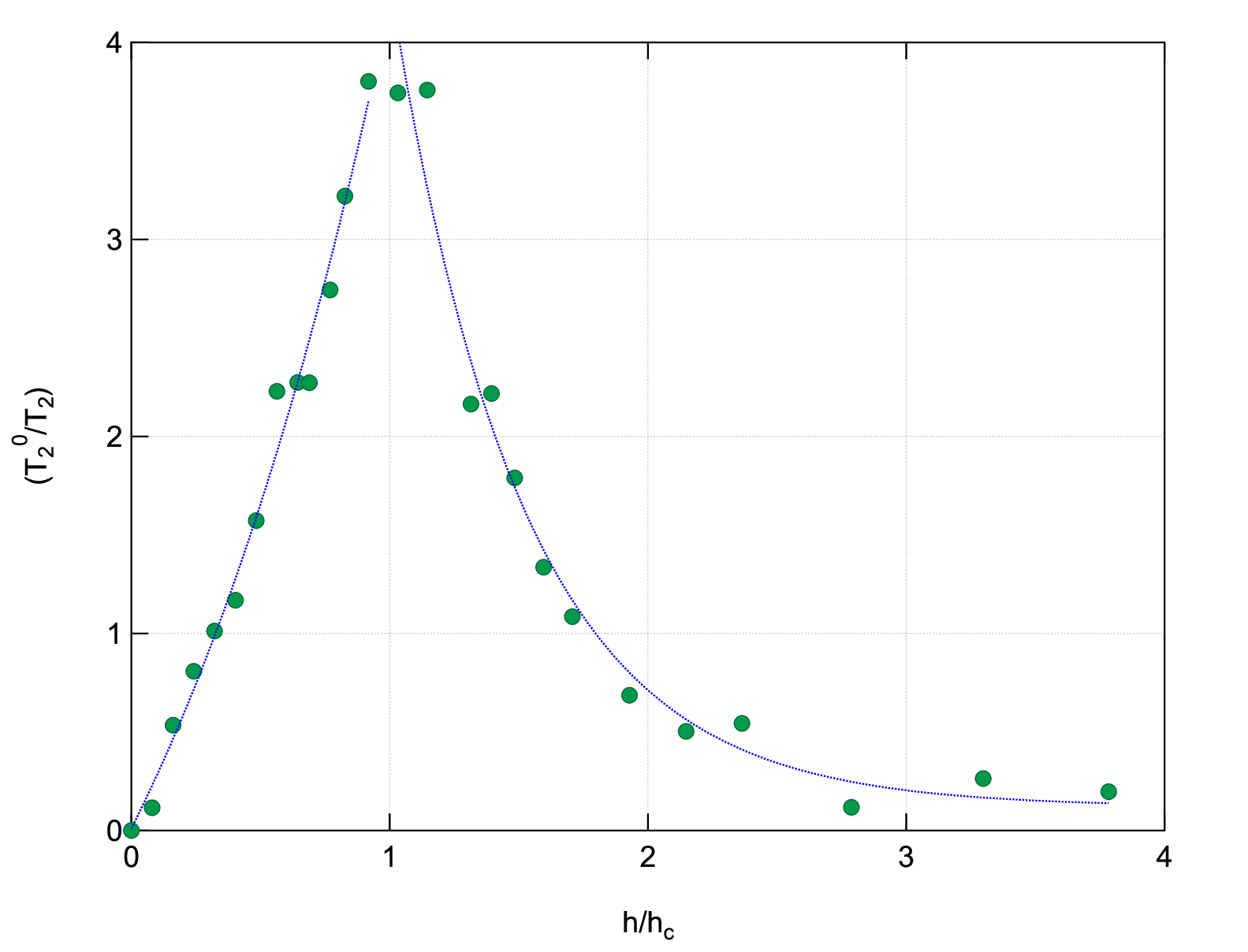}
\caption{Normalized spin decoherence rate, $T_2^0/T_2  = -\log(I(h)/I(0))$ measured in TmVO$_4$ at $T=0.77$ \cite{Nian2024}.  Here $I$ is the integrated spectral intensity measured by spin echoes, normalized such that $I(0)$ is unity, and the constant $T_2^0 \approx 6\times 10^{-5}$ sec. The sharp rise at the quantum critical field is indicative of quantum criticality.
\label{T2}
}
\end{figure}

Our study provides a resolution of the different behaviors observed in the thermodynamic and NMR spin-
echo measurements in the thulium vanadate materials. The phase diagram and heat capacity jump of the model
closely follow mean-field theory. These are well reproduced by the infinite-range model.
However, unlike mean-field theory, NMR spin-echo measurements show persistent low energy quantum critical fluctuations near the critical field. As an example, the divergent spin-spin relaxation rate is shown in Fig. 12. The infinite range transverse-field Ising model shows that both the mean-field thermodynamic and low energy
spectral behaviors can be simultaneously present in a long-ranged system.

In Fig.~13, we show the spectral function associated with the total transverse spin operator. In these finite systems, the spectral weights consist of a series of delta-function peaks that have been broadened
by a Lorentzian with a broadening parameter $\eta=10^{-5}$. The spectral functions are shown on a logarithmic scale. One can see that away from the quantum critical point at $h_c=1$, there is a gap in the spectra of order $J$ which is almost size
independent. This gap is too large to cause any low frequency NMR relaxation. In contrast, at the critical point $h_c=1$ a whole series of states are converging towards zero energy. It is these states at
low energies, characteristic of a quantum critical point, that can cause a rapid decoherence of the nuclear spins.

\begin{figure}
\begin{center}
\includegraphics[width=0.98\columnwidth]{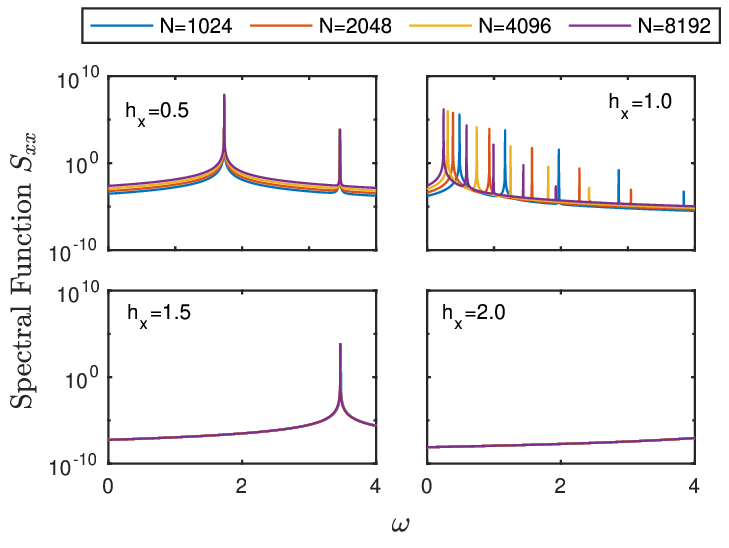}
\caption{Spectral weights associated with the total transverse spin with delta functions broadened by a Lorentzian of width $\eta=10^{-5}$ for transverse-field values of $0.5$, $1.0$,
$1.5$ and $2.0$. Away from the quantum critical point at $h_x=h_c=1$, the system has a large gap. At the quantum critical point, a number of states converge towards zero energy with increase in system size.
\label{Spectral-func}}
\end{center}
\end{figure}

The main effect of temperature in the model is that the probability of finding the system in a spin-sector $s_t$ changes as a function of temperature. Probability distribution for finding the system in different spin-sectors, for two different values of the transverse-field, for a system of $N=1024$ spins are shown in Fig.~14 for temperatures up to twice the zero-field transition temperature.  At low temperatures (up to about $T=0.2$) this probability is very sharply peaked near the highest spin sector, where the ground state resides. Even at twice the ordering temperature the probability is still peaked in a spin sector that is a fraction of the maximum spin, hence extensive in number of
spins. It would take an unphysically large temperature for the probability to finally shift to sectors with spin of order one. One can also see in the plot that the largest change in the probability distribution occurs as one crosses the phase transition.

In each such sector with an extensive spin
value, the system supports low energy quantum dynamics near the critical field. That is, the gaps go to zero in every large spin sector at $h_c$. As the main role of temperature is to shift the probability distribution to reduced $s_t$ values, low energy weight remain at the critical field even as temperature increases.  This tells us that for a sufficiently large system the quantum critical fluctuations can persist at the quantum critical point to temperatures of order the exchange constant. We note that it has been shown in literature that in a one-dimensional transverse-field Ising model, quantum critical fluctuations persist up to infinite temperature \cite{Chen2013,Quan2006}.

\begin{figure}
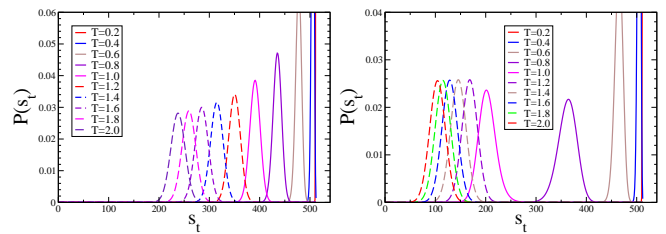

\begin{center}
\includegraphics[width=0.49\columnwidth]{wt-N1024-h1.eps}
\includegraphics[width=0.49\columnwidth]{wt-N1024-hp4.eps}
\caption{
Probability distribution for finding the system in spin-sector $S$ at the critical field $h_c=1$ (left) and deep inside the ordered phase $h_c=0.4$ (right) at several temperatures for a system of $N=1024$ spins with maximum spin of $s_t=516$. For $h<h_c$, such as $h_c=0.4$, there is a finite temperature phase transition and the largest shift in the probability distribution happens as one crosses the phase transition.
\label{Probability}}
\end{center}
\end{figure}

\section{Summary}
\label{sec:summary}

In summary, we have shown that the infinite range transverse-field Ising model exhibits quantum criticality. The energy gap goes to zero as the quantum critical field is approached from either side of the transition. The study of Quantum Fisher Information shows that the system becomes highly entangled as the quantum critical point is approached in the temperature-field plane. While the Quantum Fisher Information appears to be singular at the finite temperature phase transition, it does not serve as a witness for enhanced quantum entanglement along the finite temperature phase boundary away from the quantum critical region. In other words, the enhanced multi-partite quantum entanglement is observed only in the quantum critical region. 

Various thermodynamic quantities show deviations from mean-field behavior especially near the quantum critical field. Heat capacity has a jump at the transition which closely resembles the mean-field behavior. However, the transverse susceptibility shows a peak at $T=0$ near the quantum critical point, unlike mean-field theory. At finite temperatures also
a small but distinct peak in the transverse susceptibility remains. The spectral function associated with the transverse-susceptibility shows low energy weights only
near the quantum critical point, which persists over a range of temperatures.

This study points to a possible consistent explanation of the various observed properties
in the Thulium Vanadate materials. While the heat capacity in the material shows a jump quite similar to mean-field theory, NMR studies find quantum
critical features with a vanishing spin-gap and a rapid decoherence of nuclear spin states in the vicinity of the quantum critical field consistent with
the results of our model study. We also find peaks in
the transverse susceptibility at low temperatures in the quantum critical region and
speculate that such peaks should be present for phonon mediated transverse-field Ising systems that can be modelled with long-range interactions.
%is large but not infinite and the interactions decay slowly with
%distance as is more realistic for phonon mediated interactions and the critical behavior is mean-field, 

%\begin{figure}
%\begin{center}
%\includegraphics[angle=0,width=0.9\columnwidth]{distri.eps}
%\caption{
%Distribution
%\label{fig}}
%\end{center}
%\end{figure}

% \begin{table}[ht]
% \caption{\label{table}
% The value of $(h^T_c/J)^2$ is estimated to be
% $2.268$ in Ref.~\cite{yamamoto:87} and $2.28\pm0.03$ in Ref.~\cite{young:17b}.
% }
% \begin{tabular}{|c|.|c|.|c|}
% \hline
% Model & \multicolumn{2}{c|}{$\chiSG$} & \multicolumn{2}{c|}{$S(0)$} \\
% \hline
% & \multicolumn{1}{|c|}{$ 1/x_c$}& $ \lambda $ & \multicolumn{1}{|c|}{$1/x_c$}& $\ \gamma - 2 z \nu$\  \\
% \hline \hline
% SK           &   2.267  &  0.578    &  2.265 &   -0.509           \\
% SK (log)     & 2.268    & 0.486     &  2.265 &  -0.509            \\
% SK \cite{ye:93,huse:93,read:95,yamamoto:87}   &2.268   & $1/2$
% &2.268       & $-1/2    $ \\
% SK \cite{young:17b} &2.28(3)&  &2.28(3)&  \\
% \hline
% \end{tabular}
% \end{table}

%{\it Acknowledgments:}
%The work of RRPS is supported in part by US NSF grant
%number DMR-1306048.

\acknowledgements
We thank  I. Fisher and I. Vinograd for stimulating discussions.  This Work  was supported by the NSF under Grants No. DMR-2210613 and PHY-1852581.

\bibliography{QC.bib}

\section{Appendix}
\label{sec:appendix}
In this appendix we develop high-field perturbation theory for the infinite-range model to show that the gaps in excitation energy have finite corrections to the mean-field results even in the limit that the transverse-field $h_x$ goes to infinity. In mean-field theory, the entire paramagnetic phase is described by decoupled spins in a transverse field, for which the excitation gap corresponding to flipping a spin has energy $2h_x$. We consider the maximum $S_t$ sector of our infinite-range model in which the ground state lies for all $h_x$. In this sector, all states correspond to a uniform state analogous to a $q=0$ state in finite dimensional system.

In the ground state for $h_x$ going to infinity, all spins are pointing in the x-direction. Its energy can be calculated perturbatively and gives
\begin{equation}
    E_0= -Nh_x -\frac{J^2}{16 h_x},
\end{equation}
as $h_x\to\infty$ corrections to mean-field result goes to zero.
The one-particle state or single spin-flip state is an equal superposition of spin-flip states at each site. Its energy already changes in first order perturbation theory due to matrix elements connecting flipped spins at different sites leading to
\begin{equation}
    E_1=-(N-2)h_x -J.
\end{equation}
Similarly, the energy of two-spin flip state has energy
\begin{equation}
    E_2=-(N-4)h_x -2J.
\end{equation}
This shows that as $h_x$ goes to infinity, there remains a correction to energy gaps from the mean-field result. The one-particle excitation has a gap of $2h_x-J$ and two-particle gap has excitation gap $4h_x-2J$.
\end{document}